\documentstyle[aps,graphicx,epsf,prb,twocolumn,floats]{revtex}

\begin{document}


\title{Ferromagnetic-spin glass transition in four-dimensional
random-bond Ising model}
\author{Alexander K. Hartmann}
\address{Department of Physics, University of California,
Santa Cruz CA 95064, USA\\
      E-mail: \texttt{hartmann@bach.ucsc.edu}
        }

\date{\today}
\maketitle 
\begin{abstract}
The four-dimensional $\pm J$ random-bond Ising model is studied using
ground-state calculations. System sizes up to $N=6^4$ spins are
considered. Here it is found that the ferromagnetic-spin glass transition
occurs at a critical concentration $p_c=0.28(1)$ of the 
antiferromagnetic bonds,
which is comparable to  values found previously by high-temperature
series expansions.
The transition is characterized by a correlation-length exponent
$\nu=1.0(1)$ and an
order-parameter exponent 
$\beta=0.4(1)$. Thus, this transition is in a
different universality class from four-dimensional bond percolation,
where $\nu=0.678(50)$ and $\beta=0.639(20)$.

{\bf Keywords (PACS-codes)}: Spin glasses and other random models (75.10.Nr), 
Numerical simulation studies (75.40.Mg),
General mathematical systems (02.10.Jf).
\end{abstract}

\section*{Introduction}

The study of how order arises in nature is 
in the center of interest in
several areas of physics, especially in thermodynamics and 
statistical physics. A special focus is on the question, 
how order can emerge, even in or due to the presence of quenched disorder.
One type of system, which has attracted a lot of attention during the
last decades, are spin glasses\cite{reviewSG}.

In this work, the transition
from one ordered phase, the ferromagnetic phase, to another ordered phase,
the spin-glass phase, is studied for a four-dimensional
random-bond Ising model.  
The model is studied by means of ground-state calculations,
using the {\em genetic cluster-exact approximation (CEA)} method.
This approach has the advantage that one does not
encounter ergodicity problems or critical
slowing down like when using algorithms which are based
on Monte-Carlo methods. 
To the author's knowledge, there are no numerical studies of 
the four-dimensional random-bond Ising model (except at $p=0.5$).
However, for two\cite{bendisch92,bendisch94,kawashima97} 
and three\cite{alex-threshold} 
dimensions, the random-bond Ising model
has been studied numerically at $T=0$ already, resulting in critical values
$p_c^{\rm sq}=0.10$ for square lattices respectively
$p_c^{cub}=0.22$ for cubic lattices.

The model treated here consists of 
$N=L^4$ Ising spins $\sigma_i=\pm 1$ on a hypercubic lattice.
The Hamiltonian of this system is given by
\begin{equation}
H \equiv - \sum_{\langle i,j\rangle} J_{ij} \sigma_i \sigma_j
\end{equation}
where $\langle \ldots \rangle $ denotes a sum over pair of nearest neighbors.
Systems with quenched disorder of the bonds $J_{ij}=\pm 1$ are studied.
The ferromagnetic-spin glass transition is driven by increasing the fraction
$p$ of antiferromagnetic ($J_{ij}=-1$) (AF) 
bonds. For low concentration $p$, the system is
ferromagnetically ordered, while for intermediate values of $p$,
spin-glass order arises. For very high concentration, the system  shows 
AF ordering.

\begin{figure}[htb]
\begin{center}
\epsfxsize=\columnwidth
\epsfbox{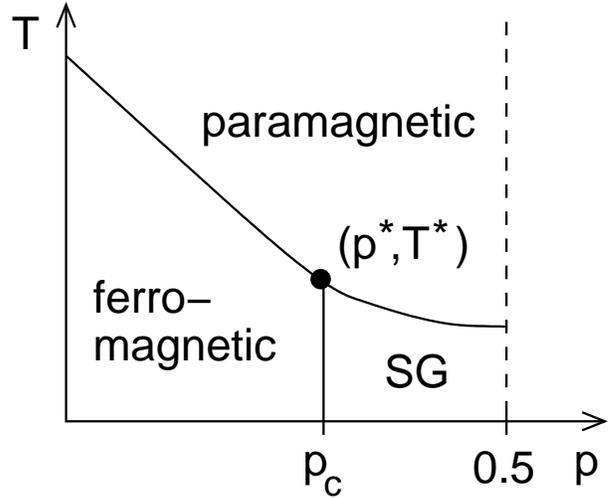}
\end{center}
\caption{Schematic phase diagram of the four-dimensional random-bond model.
The diagram is symmetric with respect to $p=0.5$, with an
antiferromagnetically ordered phase for large $p$. SG denotes the
spin glass phase. The multicritical point is denoted by $(p^{*},T^{*})$.}
\label{figPhaseDiagram}
\end{figure}

In previous work, the model has been studied using high-temperature
series expansions\cite{rajan1983,fisch1991,singh1996}. In Fig.
\ref{figPhaseDiagram} the schematic phase diagram is presented. 
The line connecting  the multicritical point $(p^{*},T^{*})$ with
$(p=p_c,T=0)$ starts 
vertically\cite{nishimori1981,ledoussal1988} at the multicritical point.
Therefore, $p_c\approx p^{*}$ can be expected.
The values previously obtained for the concentration
at the multicritical point are 
$p^{*}\approx 0.3$ (Ref. \onlinecite{rajan1983}),
 $p^{*}=0.28(1)$  (Ref. \onlinecite{fisch1991}) and
 $p^{*}=0.290(1)$  (Ref. \onlinecite{singh1996}).

The rest of the paper is organized as follows: Next, the algorithm
to generate  spin-glass ground states in four dimensions 
is briefly explained.  Then, the results describing the
ferromagnetic-spin glass transition
 are presented and finally a summary is given.

\section*{algorithm}

From the computational point of
view, the calculation of spin-glass ground states is very demanding,
because it belongs to the class of the NP-hard problems
\cite{barahona1982,opt-phys2001}. 
This means that only algorithms are available,
for which the running time on a computer  increases
exponentially with the system size. 
Only for the special case of a planar system without magnetic field, e.g.
a square lattice with periodic boundary conditions in at most one direction,
exist efficient polynomial-time ``matching'' algorithms\cite{bieche1980}.

For higher dimensions, one must rely on exact methods like
branch-and-bound \cite{hartwig84,klotz} or 
branch-and-cut\cite{simone95,simone96}, which are able to treat 
only small systems.
For that reason, recently some heuristics have been introduced. By using a
hierarchical approach\cite{houdayer1999} one can calculate true
ground states in four dimensions up to size $L=5$, while for $L=6$ it is
not clear whether true ground states were found\cite{houdayer2001}.

The method applied here\cite{alex-stiff4d}, 
is able to calculate ground-states in four
dimensions up to size $L=7$. The technique is based on a special genetic
algorithm \cite{pal96,michal92} and on cluster-exact approximation
\cite{alex2} which is  an optimization method designed especially
for spin glasses. Now a brief description of the techniques are given.

Genetic algorithms are biologically motivated. An optimal
solution is found by treating many instances of the problem in
parallel, keeping only better instances and replacing bad ones by new
ones (survival of the fittest).
The genetic algorithm starts with an initial population of $M_i$
randomly initialized spin configurations (= {\em individuals}),
which are linearly arranged in
a ring. Then $\nu \times M_i$ times two neighbors from the population
are taken (called {\em parents}) and two {\em offspring} are created
using the so called triadic crossover \cite{pal95}.
Then a mutation with a rate of $p_m$
is applied to each offspring, i.e. a fraction $p_m$ of the
spins is reversed.
Next, for both offspring the energy is reduced by applying the
CEA. This algorithm 
constructs iteratively and randomly
a non-frustrated cluster of spins, whereas
spins with many unsatisfied bonds are more likely to be added to the
cluster.
The  non-cluster spins act like local magnetic fields on the cluster spins.
For the spins of the cluster, an energetic minimum state can be
calculated in polynomial time
by using graph-theoretical methods
\cite{claibo,knoedel,swamy}: an equivalent network is constructed
\cite{picard1}, the maximum flow is calculated
\cite{traeff,tarjan} and the spins of the
cluster are set to the orientations leading to a minimum in energy.
This minimization step
is performed $n_{\min}$ times for each offspring.

Afterwards each offspring is compared with one of its parents. The
pairs are chosen in the way that the sum of the phenotypic differences
between them is minimal. The phenotypic difference is defined here as the
number of spins  where the two configurations differ. Each
parent is replaced if its energy is not lower (i.e. better) than the
corresponding offspring.

After this creation of offspring has been performed
 $\nu \times M_i$ times, the population
is halved: From each pair of neighbors the configuration
 which has the higher energy is eliminated. If not more than 4
individuals remain, the process is stopped and the best individual
is taken as result of the calculation.

The whole algorithm is performed $n_R$ times and all final configurations
which exhibit the lowest energy are stored, resulting in $n_g$ statistical
independent ground-state configurations. By comparison with an
exact method, it was shown\cite{alex-stiff3d} that genetic CEA, 
with an appropriate choice of the parameters
($M_i$, $\nu$, $n_{\min}$, $p_m$), indeed calculates true ground states.

The probability that a certain ground-state configuration
is found by this method is
not equal for all ground states \cite{alex-false}, i.e. the algorithm
imposes a bias. In this work, the
magnetization is the main quantity of interest.
To test, how large the influence of this bias is, for $L=6$, $p=0.27$,
where fluctuations of the magnetization are the largest,  64
realizations were considered. For each realization, (biased) ground states
were generated using 200 independent runs. Next, for each realization
a (thermodynamically correct)
 set of ground states was generated in such a way that each
configuration contributes with the same probability\cite{alex-equi}. 
For both sets
of ground states the magnetization was evaluated. For the biased set
an average value of $m=0.501(11)$ was obtained, while the correct
thermodynamic result is $m=0.504(12)$. This shows that the influence
of the bias on the result of the magnetization is small. Please
note that that is  in contrast
to other quantities like the distribution $P(q)$ of overlaps:
The overlap  $q^{\alpha\beta}$
between two independent ground states $\{\sigma^{\alpha}_i\}$,
$\{\sigma^{\beta}_i\}$ is given by 
$q=\frac{1}{N}\sum_i\sigma^{\alpha}_i\sigma^{\beta}_i$. When measuring
the fraction $x(0.5)=\int_{-0.5}^{0.5} P(q)dq$ of small overlaps,
one obtains an average value $x(0.5)=0.051(7)$ for the biased set 
of states, while the from the correct thermodynamic treatment one
obtains quite a different result $0.014(8)$.

Hence, when restricting the measurement to quantities depending solely on the
magnetization,
it is sufficient to use the biased data. This allows much 
shorter running times,
because only few ground states per realization are needed.

\section*{Results}

With the genetic cluster-exact approximation, it it possible\cite{alex-stiff4d}
 to obtain ground-states for four-dimensional systems up to $L=7$. Since
the study of the transition involves the calculation of a Binder
cumulant, it is necessary to average over many samples of the disorder.
For that reason, only sizes $L\le 6$ were considered in this work.
A mutation rate of $p_m=0.05$ was used.
The other simulation parameters are shown in Tab. I.

\begin{table}[htb]
\begin{center}
\begin{tabular}{cccccc}
\hline
$L$ & $M_i$ & $\nu$ & $n_{\min}$ & $\tau$ (sec) & $N_L$ \\ \hline
3 & 16 & 4 & 4 & 3 & $10^5$ \\
4 & 16 & 4 & 4 & 14 & 15000 \\
5 & 256 & 6 & 10 & 4800 & 13000 \\
6 & 256 & 6 & 10 & 7300 & 20500 \\
7 & 512 & 12 & 20 & 14000 & -
\end{tabular}
\end{center}
\caption{Simulation parameters: $L$ = system size, $M_i$ = initial size of
population, $\nu$ = average number of offspring per configuration, $n_{\min}$
= number of CEA minimization steps per offspring, $\tau$ = typical computer
time per ground state on a 80MHz PPC601, $N_L$ = total number of realizations
of the random variables.}
\label{tab_parameters}
\end{table}

\begin{figure}[htb]
\begin{center}
\epsfxsize=\columnwidth
\epsfbox{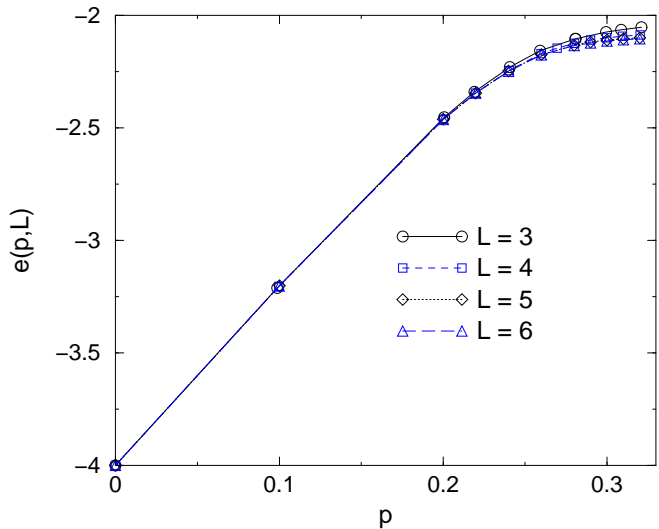}
\end{center}
\caption{Ground-state energy $e(p,L)$ 
 as a function of the AF-bond concentration $p$ for system sizes 
$L=3,4,5,6$. Error bars are much smaller than system sizes. Lines are
guides for the eyes only.}
\label{figEnergy}
\end{figure}

Systems with concentrations of the antiferromagnetic bonds in the range
$p \in [0,0.32]$ were treated. For each realization $n_R=5$ independent runs
were performed. The resulting ground-state energy as
a function of the parameter $p$ is shown in Fig.
\ref{figEnergy} for different system sizes. For small concentrations $p$,
the ground state is mainly ferromagnetic.
It follows that all the AF bonds are not satisfied, so the ground state
energy increases linearly with $p$ like $e(p)\approx -4+8p$.
For larger concentrations, the
ground-state energy approaches the $p=0.5$ limit.
With increasing $L$ the ground state energy decreases,
because the periodic boundary conditions impose less
constraints on the system. For $L\to\infty$ and $p=0.5$ a ground
state energy of $e_{\infty}(0.5)=-2.095(1)$ was found\cite{alex-stiff4d}.

\begin{figure}[htb]
\begin{center}
\epsfxsize=\columnwidth
\epsfbox{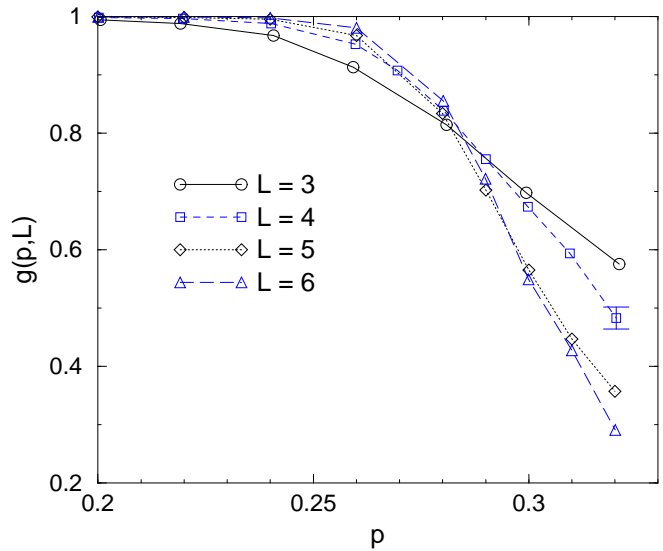}
\end{center}
\caption{Binder cumulant $g(p,L)$ 
of the ground-state magnetization as a function of the
AF-bond concentration $p$ for system sizes $L=3,4,5,6$. For clarity,
only the largest error bar occurring in the data is shown. Lines are
guides for the eyes only.}
\label{figBinder}
\end{figure}

From Fig. \ref{figEnergy} is clear that the
 energy as function of concentration is not a suitable quantity
for determining the critical concentration $p_c$,
where the ferromagnetic behavior disappears.
For this purpose the Binder cumulant \cite{binder81,bhatt85}
\begin{equation}
g(p,L)\equiv\frac{1}{2}
\left( 3-\frac{[\langle M^4\rangle]_J }{[\langle M^2\rangle]_J^2}\right)
\end{equation}
for the magnetization $M\equiv\frac{1}{N}\sum_i \sigma_i$ is evaluated.
The average $\langle \ldots \rangle$ denotes the average over
different ground states of a realization, while $[\ldots ]_J$
is the average over the disorder.
In Fig. \ref{figBinder} the Binder cumulant is shown for $L=3,4,5,6$.
To keep the figure clear, only the largest error bar 
is shown. All curves intersect near $p_c=0.28\pm 0.01$. Because of
the huge numerical effort, the simulations are restricted to small system
sizes, which prevents a more accurate result.

Only few studies concerning the four-dimensional random-bond model
have been performed before. Using high-temperature series 
expansions for 
the location of the multicritical point $(p^{*},T^{*})$, 
values of $p^{*}\approx 0.3$ (Ref. \onlinecite{rajan1983}),
 $p^{*}=0.28(1)$  (Ref. \onlinecite{fisch1991}) and
 $p^{*}=0.290(1)$  (Ref. \onlinecite{singh1996}) were determined.
These results compare well with the value of $p_c$ 
obtained here, indicating that the
the line connecting $(p_c,0)$ with $(p^{*},T^{*})$ is vertical, or nearly so,
as expected\cite{nishimori1981,ledoussal1988}.

\begin{figure}[htb]
\begin{center}
\epsfxsize=\columnwidth
\epsfbox{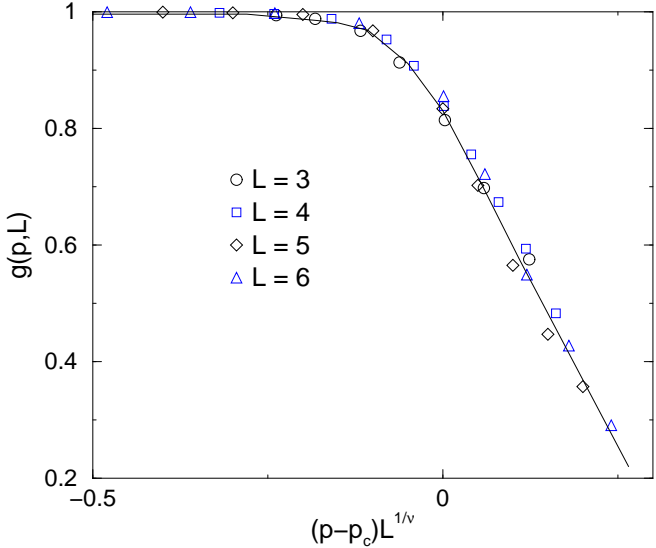}
\end{center}
\caption{Scaling plot of the Binder cumulant as a function of
$(p-p_c)L^{1/\nu}$ with $p_c=0.28$ and $\nu=1.0$. 
The line is a guide for the eyes only.}
\label{figBinderScale}
\end{figure}

For the Binder cumulant the following finite-size scaling relation
is assumed \cite{bhatt85}

\begin{equation}
g(p,L)=\tilde{g}(L^{1/\nu}(p-p_c))\,.
\end{equation}
By plotting $g(p,L)$ against $L^{1/\nu}(p-p_c)$ with correct
parameters $p_c,\nu$ the datapoints for different system sizes should
collapse onto a single curve near $(p-p_c)=0$. 
The best results were obtained for
$p_c=0.28$ and $1/\nu=1.0$. In Fig. \ref{figBinderScale} the resulting
scaling plot is shown. It is possible to change the value of $\nu$ in
a wide range without large effects on the scaling plot. Thus, here a value of
$\nu=1.0(2)$ is estimated. 
This is compatible with the value $\nu=0.8(1)$ found by a high-temperature
series expansion\cite{singh1996} at the multicritical 
point.
Although there is no direct correspondence between the $T=0$
ferromagnetic-spin glass transition and the finite temperature
($p=0.5$) paramagnetic-spin glass transition at $T_c=2.03(3)$, it is remarkable
that by Monte-Carlo simulations\cite{marinari1999} of four-dimensional spin
glasses near $T_c$ a similar value of $\nu=1.0(1)$ was found.
Finally, it should be pointed out, that the result given above is
not compatible with the corresponding exponent for four-dimensional
bond percolation\cite{adler1990} of $\nu=0.678(50)$.

\begin{figure}[htb]
\begin{center}
\epsfxsize=\columnwidth
\epsfbox{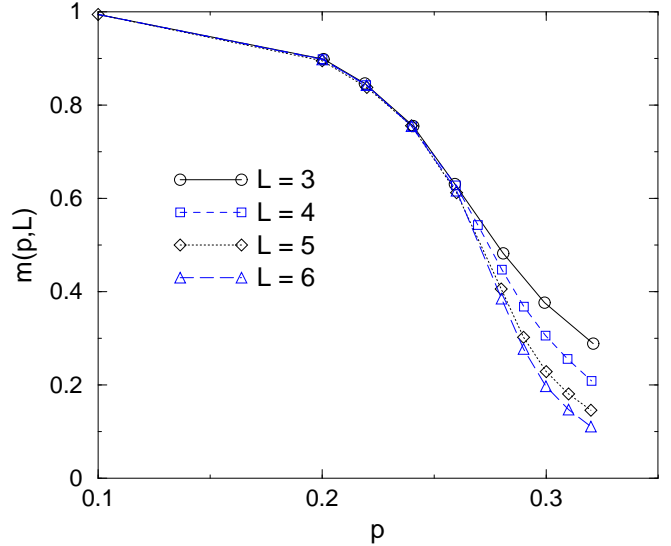}
\end{center}
\caption{Average ground-state magnetization $m(p,L)$ as a function of
the AF-bond concentration $p$ for system sizes $L=3,4,5,6$. All error
bars are much smaller than symbol size. Lines are
guides for the eyes only.}
\label{figMag}
\end{figure}

\begin{figure}[htb]
\begin{center}
\epsfxsize=\columnwidth
\epsfbox{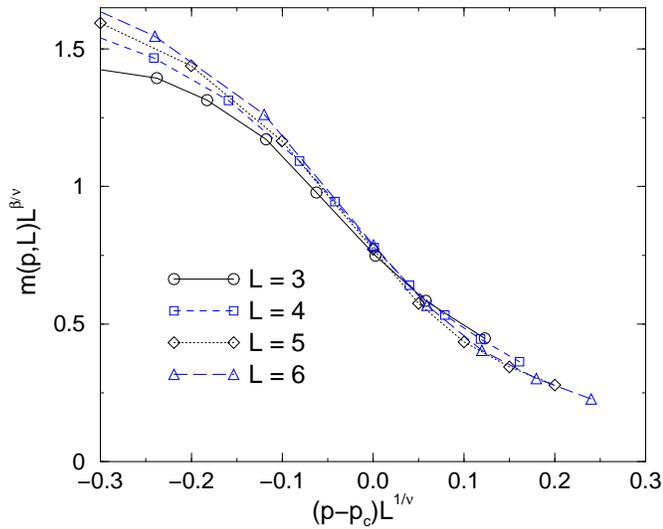}
\end{center}
\caption{Scaling plot of the rescaled magnetization
  $m(p,L)L^{\beta/\nu}$
as a function of $(p-p_c)L^{1/\nu}$ with $p_c=0.28$, 
$\nu=1.0$ and $\beta=0.4$. 
Lines are guides for the eyes only.}
\label{figMagScale}
\end{figure}

In Fig. \ref{figMag} the behavior of the average magnetization
$m\equiv [ \langle M \rangle ]_J$ is shown as a function of $p$
for different system sizes.
This quantity has 
the standard finite-size  scaling form \cite{binder_heermann}
\begin{equation}
m(p,L)=L^{-\beta/\nu}\tilde{m}(L^{1/\nu}(p-p_c))\,.
\end{equation}
By plotting $L^{\beta/\nu}m(p,L)$ against $L^{1/\nu}(p-p_c)$ with correct
parameters $p_c,\beta,\nu$ the datapoints for different system sizes should
collapse onto a single curve near $(p-p_c)=0$.
The best result was obtained using $p_c=0.28$, 
$1/\nu=1.0$ and $\beta/\nu=0.4$.
It is shown in Fig. \ref{figMagScale} for $L=3,4,5,6$. 
From variations of the value $\beta/\nu$, the
value of the exponent $\beta=0.4(1)$ is estimated. Again, this values
 significantly differs from the exponent $\beta=0.693(20)$ 
found\cite{adler1990}
for the order parameter of the four-dimensional bond-percolation.

\section*{Summary}

The ferromagnetic-spin glass transition of the four-dimensional
 random-bond Ising model was studied at $T=0$ using ground-state
calculations. The genetic cluster-exact approximation
method was applied. Because of the high computational effort,
only small systems of size $N\le 6^4$ could be studied. Since the
ground-state problem is NP-hard, it is very unlikely that significantly
larger sizes can be studied in the near future.

By evaluating the Binder cumulant, a critical concentration of
$p_c=0.28(1)$ was found, which is comparable to values which were
obtained by  high-temperature series expansions.

Using a finite-size scaling analysis, the critical exponents for the
divergence of the correlation length $\nu=1.0(1)$ and for the
magnetization $\beta=0.4(1)$ were determined. Hence, the transition is
clearly in a different universality class than four-dimensional bond
percolation.

\section{Acknowledgements}

The author thanks A.P. Young for critically reading the manuscript 
and various other support and R. Fisch for helpful communications.
The work was supported by the Graduiertenkolleg
``Modellierung und Wissenschaftliches Rechnen in 
Mathematik und Naturwissenschaften'' at the
{ In\-ter\-diszi\-pli\-n\"a\-res Zentrum f\"ur Wissenschaftliches
  Rechnen} in Heidelberg and the
{ Paderborn Center for Parallel Computing}
 by the allocation of computer time.
The author acknowledges financial support from the DFG (Deutsche 
Forschungsgemeinschaft)
under grant Ha 3169/1-1.

\end{document}